\documentclass{SciPost}

\hypersetup{
    colorlinks,
    linkcolor={red!50!black},
    citecolor={blue!50!black},
    urlcolor={blue!80!black}
}

\usepackage[bitstream-charter]{mathdesign}
\DeclareSymbolFont{usualmathcal}{OMS}{cmsy}{m}{n}
\DeclareSymbolFontAlphabet{\mathcal}{usualmathcal}

\fancypagestyle{SPstyle}{
\fancyhf{}
\lhead{\colorbox{scipostblue}{\bf \color{white} ~SciPost Physics Lecture Notes }}
\rhead{{\bf \color{scipostdeepblue} ~Submission }}

\fancyfoot[C]{\textbf{\thepage}}
}

\usepackage{bm}
\newcommand{\figref}[2]{%
  \hyperref[#1]{Fig.~\ref{#1}#2}%
}
\newcommand{\secref}[1]{%
  \hyperref[#1]{Sec.~\ref{#1}}%
}
\newcommand{\eref}[1]{%
  \hyperref[#1]{Eq.~\ref{#1}}%
}
\newcommand{\tavg}[1]{\left\langle #1 \right\rangle}
\usepackage{physics}

\begin{document}

\pagestyle{SPstyle}

\begin{center}{\Large \textbf{\color{scipostdeepblue}{
Simplified derivations for high-dimensional convex\\learning problems
}}}\end{center}

\begin{center}\textbf{
David G. Clark\textsuperscript{1,2$\star$} and
Haim Sompolinsky\textsuperscript{3,4,5$\dagger$}
}\end{center}

\begin{center}
{\bf 1} Zuckerman Institute, Columbia University, New York, NY, USA
\\
{\bf 2}
Kavli Institute for Brain Science, Columbia University, New York, NY, USA
\\
{\bf 3} Center for Brain Science, Harvard University, Cambridge, MA, USA
\\
{\bf 4} Program in Neuroscience, Harvard Medical School, Boston, MA, USA
\\
{\bf 5} Edmond and Lily Safra Center for Brain Sciences, Hebrew University, Jerusalem, Israel
\\[\baselineskip]
$\star$ \href{mailto:dgc2138@cumc.columbia.edu}{\small dgc2138@cumc.columbia.edu}\,,\quad
$\dagger$ \href{mailto:hsompolinsky@mcb.harvard.edu}{\small hsompolinsky@mcb.harvard.edu}
\end{center}

\section*{\color{scipostdeepblue}{Abstract}}
\textbf{\boldmath{%
Statistical-physics calculations in machine learning and theoretical neuroscience often involve lengthy derivations that obscure physical interpretation. Here, we give concise, non-replica derivations of several key results and highlight their underlying similarities. In particular, using a cavity approach, we analyze three high-dimensional learning problems: perceptron classification of points, perceptron classification of manifolds, and kernel ridge regression. These problems share a common structure---a bipartite system of interacting feature and datum variables---enabling a unified analysis. Furthermore, for perceptron-capacity problems, we identify a symmetry that allows derivation of correct capacities through a naïve method.
}}

\vspace{\baselineskip}

\noindent\textcolor{white!90!black}{%
\fbox{\parbox{0.975\linewidth}{%
\textcolor{white!40!black}{\begin{tabular}{lr}%
  \begin{minipage}{0.6\textwidth}%
    {\small Copyright attribution to authors. \newline
    This work is a submission to SciPost Physics Lecture Notes. \newline
    License information to appear upon publication. \newline
    Publication information to appear upon publication.}
  \end{minipage} & \begin{minipage}{0.4\textwidth}
    {\small Received Date \newline Accepted Date \newline Published Date}%
  \end{minipage}
\end{tabular}}
}}
}


\vspace{10pt}
\noindent\rule{\textwidth}{1pt}
\tableofcontents
\noindent\rule{\textwidth}{1pt}
\vspace{10pt}


\section{Introduction}

Tools from statistical physics have elucidated important properties of artificial and biological neural networks \cite{advani2013statistical, carleo2019machine, bahri2020statistical} such as storage-capacity limits \cite{amit1985storing, gardner1988space, mezard1989space, chung2016linear, chung2018classification}, generalization characteristics \cite{bordelon2020spectrum, canatar2021spectral, li2024representations}, and network dynamics \cite{sompolinsky1988chaos, van1996chaos, clark2024connectivity}. The replica method has been particularly valuable for deriving exact results in learning problems where both the data dimension and sample size approach infinity with a fixed ratio. However, replica calculations often involve complex derivations that can obscure the underlying physics and potentially limit their adoption in the machine-learning and theoretical-neuroscience communities. 

Here, we show that derivations for several high-dimensional convex learning problems can be substantially simplified and unified. Specifically, we give concise derivations using the cavity method, which offers a more intuitive alternative to the replica method. In each section's \textit{Background}, we contextualize these calculations within existing literature, including previous cavity approaches. Notably, we focus on solutions of optimality conditions given random data rather than computing full partition functions or solution-space volumes. These derivations should make generalization to variants of these problems easier. This approach, sometimes called the zero-temperature cavity method \cite{mezard2003cavity}, has been applied previously, including to problems with a bipartite structure \cite{ramezanali2015cavity, rocks2022memorizing}, which will be of interest to us (see below). 

We illustrate this unified cavity approach through three distinct problems. First, we revisit Gardner's calculation of perceptron capacity \cite{gardner1988space}, the maximum number of randomly chosen points that can be linearly separated with a fixed margin. This seminal result connected statistical mechanics to learning theory and laid a mathematical foundation for analyzing linear separability in high-dimensional spaces. Second, we investigate perceptron classification of manifolds \cite{chung2016linear, chung2018classification}, where each input comprises a structured manifold of points rather than a single point. This extension captures both the geometric structure of real data and the fact that the outputs of certain neural circuits and deep-learning systems must be invariant to such structure \cite{cohen2020separability}. Finally, we analyze kernel ridge regression, a supervised learning method that optimizes the balance between prediction accuracy and function complexity through regularization \cite{bordelon2020spectrum, canatar2021spectral}. This analysis provides insights into deep-learning systems through connections between neural networks and kernel methods \cite{jacot2018neural}.

While these problems appear distinct---involving points versus manifolds, classification versus regression, and so on---we show that they share a bipartite structure of interacting feature and datum variables, enabling a unified analysis (\figref{fig:cavity_fig}). Moreover, for perceptron-capacity problems, we identify a symmetry that enables the derivation of correct capacities through a naïve method requiring only a couple of lines of algebra.

These derivations have natural extensions to account for correlation structure in data, e.g., patterns with correlated features or manifolds with correlated orientations. The bipartite structure we consider is also exemplified by architectures including Hopfield networks and their ``modern'' counterparts. This approach can also be applied to time-dependent problems to derive dynamical mean-field theory equations. We describe these extensions in the \textit{Conclusion}.

\section{Classic Gardner problem}

\subsection{Background}

\begin{figure*}
    \centering
    \includegraphics[width=\textwidth]{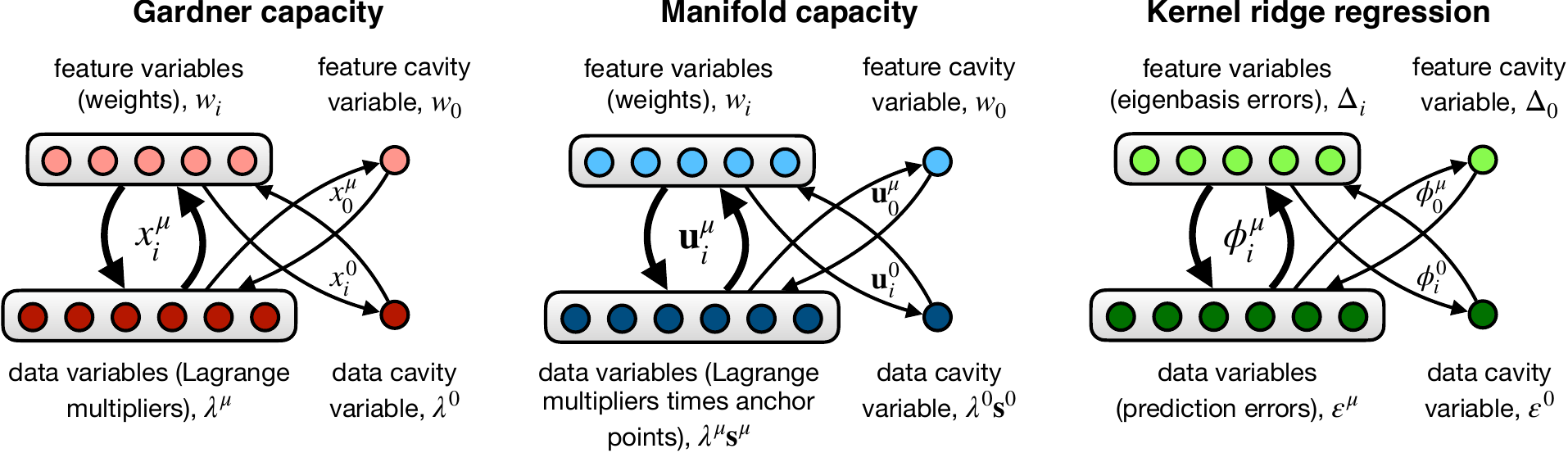}
    \caption{Bipartite structure and cavity analysis of three high-dimensional learning problems. Each panel illustrates the interaction between feature variables (top) and datum variables (bottom), connected by random couplings (arrows). Left: Gardner capacity problem with weight variables $w_i$ interacting with Lagrange multipliers $\lambda^\mu$ through random data components $x_i^\mu$. Middle: manifold capacity problem with weights $w_i$ interacting with products of Lagrange multipliers and anchor points (``vector-Lagrange multipliers'') $\lambda^\mu\bm{s}^\mu$ through random $(D+1)$-dimensional embedding vectors $\bm{u}_i^\mu$. Right: kernel ridge regression with eigenbasis errors $\Delta_i$ interacting with prediction errors $\varepsilon^\mu$ through random eigenfunction components $\phi_i^\mu$. In each case, the cavity method introduces new variables (right side of each panel) while analyzing perturbations to the existing unperturbed system (left side). Thick arrows indicate the primary couplings in the unperturbed system and thin arrows show the additional couplings introduced for the cavity variables.}
\label{fig:cavity_fig}
\end{figure*}

The capacity of a neural system refers to the maximum amount of information it can reliably store, process, or recall through configuration of its internal degrees of freedom, e.g., synaptic weights. The capacity of a perceptron for classifying random points with a margin was first calculated by Gardner \cite{gardner1988space}; the zero-margin case was presented by Cover two decades prior \cite{cover1965geometrical}. Here ``capacity'' refers to the number of randomly sampled $N$-dimensional points, with random binary labels, that can be linearly classified with high probability (we state the problem precisely below). Gardner's insight was to analyze the volume of normalized weight vectors capable of correctly separating the random points. For large $N$, the normalized logarithm of this volume takes a consistent, order-one value across data realizations, i.e., it is \textit{self-averaging}. Using the replica method, Gardner computed the average of this log-volume over the data distribution to determine the critical capacity as a function of margin.

Further work by Gardner used an energy function counting the number of violations of the margin constraint \cite{gardner1988optimal}. This approach used the replica method to analyze the data-averaged log-partition function, enabling calculation of ground-state properties such as the minimum number of classification errors for a given margin and number of data. M{\'e}zard \cite{mezard1989space} then showed how to compute this partition function using a cavity approach. Most recently, Agoritsas et al. \cite{agoritsas2018out} presented a cavity method for the dynamics of perceptron learning.

The calculation presented here uses a cavity method, but differs from \cite{mezard1989space} in two ways. First, we analyze solutions of the Karush-Kuhn-Tucker (KKT) optimality conditions rather than computing a full partition function that would allow for analysis of classifier errors and finite-temperature effects. Second, we explicitly introduce a bipartite structure involving interacting feature and datum variables (weights and Lagrange multipliers, respectively). Both of these aspects simplify the calculation, and the latter reveals shared structure with other problems.

\subsection{Problem formulation}

The data in the problem comprise $P$ points, each represented by an $N$-dimensional vector with components ${x}^\mu_i$, where $i = 1,\ldots,N$ indexes vector components and $\mu = 1,\ldots,P$ indexes points. Each point has a binary label $y^\mu \in \{-1, +1\}$. The perceptron must find a weight vector with components $w_i$ that correctly classifies all points. We take this to mean that, for all $\mu$,
\begin{equation}
    \label{eq:classification_condition}
    y^\mu \sum_{i=1}^{N} w_i x^\mu_i \geq 1.
\end{equation}
For convenience, we define a local field $h^\mu$ as
\begin{equation}
    \label{eq:local_field}
    h^\mu = y^\mu \sum_{i=1}^{N} w_i x^\mu_i - 1,
\end{equation}
which lets us write the classification condition as $h^\mu \geq 0$.

A linear classifier's margin, given by the inverse weight-vector norm,
\begin{equation}
    \text{margin} = \frac{1}{\sqrt{\sum_{i=1}^{N} w_i^2}},
\end{equation}
is the distance between the homogeneous decision hyperplane, defined by $\sum_{i=1}^{N} w_i x_i = 0$, and the affine hyperplanes beyond which points are correctly classified, defined by $\sum_{i=1}^{N} w_i x_i = \pm 1$. A large margin promotes better generalization by making the classifier robust to data perturbations. Support vector machines maximize this margin (equivalently, minimize the weight-vector norm) while maintaining correct classification through the convex optimization problem
\begin{equation}
    \label{eq:optimization_problem}
    \text{minimize } \frac{1}{2}\sum_{i=1}^{N} w_i^2 \quad \text{subject to } h^\mu \geq 0 \text{ for all } \mu.
\end{equation}
The corresponding Lagrangian is
\begin{equation}
    \label{eq:lagrangian}
    \mathcal{L} = \frac{1}{2}\sum_{i=1}^{N} w_i^2 - \sum_{\mu=1}^{P} \lambda^\mu h^\mu,
\end{equation}
where $\lambda^\mu$ are Lagrange multipliers enforcing correct classification. The optimal solution satisfies the KKT conditions,
\begin{align}
    \label{eq:kkt_stationarity}
    w_i &= \sum_{\mu=1}^{P} \lambda^\mu y^\mu x_i^\mu &&\text{(stationarity, $d\mathcal{L}/d w_i = 0$)}, \\
    \label{eq:kkt_primal_feasibility}
    h^\mu &\geq 0 &&\text{(primal feasibility)}, \\
    \label{eq:kkt_dual_feasibility}
    \lambda^\mu &\geq 0 &&\text{(dual feasibility)}, \\
    \label{eq:kkt_complementary_slackness}
    \lambda^\mu h^\mu &= 0 &&\text{(complementary slackness)}.
\end{align}

To answer the Gardner capacity problem, we fix the weight-vector norm $\sqrt{\sum_{i=1}^{N} w_i^2} = \sqrt{N}/\kappa$, where $\kappa$ (technically, $\kappa/\sqrt{N}$) is the margin, and take the limit where both the input dimension $N$ and the number of points $P$ approach infinity while maintaining a fixed ratio, given by
\begin{equation}
    \alpha = \frac{P}{N}.
\end{equation}
Taking this limit while enforcing the KKT conditions yields a constraint relating $\alpha$ to $\kappa$, where $\alpha$ represents the maximum number of points, normalized by dimension, that can be linearly classified with margin $\kappa$ (i.e., the critical capacity). To generate an infinite number of points, and to obtain results that do not depend on the particular data realization, we assume a random process that generates the points, which we now describe.  

\subsection{Random-data assumption}

We assume that the data components $x^\mu_i$ are independent and identically distributed (i.i.d.). In particular, we assume first- and second-order statistics given by
\begin{align}
    \tavg{x^\mu_i} &= 0, \\
    \tavg{x^\mu_i x^\nu_j} &= \frac{1}{N} \delta^{\mu\nu} \delta_{ij}. \label{eq:data_variance}
\end{align}
Throughout, we use $\tavg{\cdots}$ to denote an average over the data distribution. Without loss of generality, we set $y^\mu = 1$ for all $\mu$\footnote{The preservation of generality is particularly obvious if the symmetry property $P(x^\mu_i) = P(-x^\mu_i)$ holds: in this case, $y^\mu x_i^\mu$ has the same distribution as $x_i^\mu$, and the problem depends on $y^\mu$ only through $y^\mu x_i^\mu$. Even without this symmetry property, no loss of generality is incurred since the data and labels are sampled independently and $(y^\mu)^2 = 1$.}. The detailed form of the single-component distribution $P(x^\mu_i)$ does not enter into the calculation due to the central limit theorem.

\subsection{Naïve mean-field analysis}

We begin with a concise ``naïve mean-field'' analysis that yields the correct capacity despite incorrect independence assumptions. Subsequently, by way of comparison to a cavity analysis, we will show that this naïve approach is guaranteed to yield the correct capacity due to a symmetry in the problem description.

Starting from the stationarity condition (\eref{eq:kkt_stationarity}), we incorrectly assume statistical independence between Lagrange multipliers $\lambda^\mu$ and data $x^\mu_i$, yielding
\begin{align}
    \label{eq:weight_normalization}
    \tavg{w_i^2} &= \sum_{\mu,\nu=1}^{P} \underset{=\delta^{\mu\nu}/N}{\underbrace{\tavg{x^\mu_i x^\nu_i}}} \tavg{\lambda^\mu \lambda^\nu} = \alpha\tavg{(\lambda^\mu)^2}.
\end{align}
From weight-vector normalization $\tavg{w_i^2} = 1/\kappa^2$, we obtain a formula for the capacity,
\begin{equation}
    \label{eq:naive_capacity}
    \alpha^{-1} = \kappa^2 \tavg{(\lambda^\mu)^2}.
\end{equation}
Next, by substituting the stationarity condition (\eref{eq:kkt_stationarity}) into the definition of the local field (\eref{eq:local_field}), we obtain
\begin{equation}
    \label{eq:local_field_expanded}
    h^\mu = \sum_{\nu=1}^P  \sum_{i=1}^{N} x^\mu_i x^\nu_i \lambda^\nu - 1.
\end{equation}
Separating diagonal ($\mu=\nu$) and off-diagonal ($\mu\neq\nu$) terms gives
\begin{equation}
    \label{eq:local_field_split}
    h^\mu = \left[ \sum_{i=1}^{N} (x^\mu_i)^2 \right] \lambda^\mu + \sum_{\nu\neq \mu} \sum_{i=1}^{N} x^\mu_i x^\nu_i \lambda^\nu - 1.
\end{equation}
The square-bracket term is unity at large $N$ due to the choice of data variance (\eref{eq:data_variance}). Under the incorrect independence assumption, the second term, which we denote by ${\eta^\mu = \sum_{\nu\neq \mu} \sum_{i=1}^{N} x^\mu_i x^\nu_i \lambda^\nu}$, is Gaussian (by the central limit theorem) with
\begin{align}
    \label{eq:t_mu_mean}
    \tavg{\eta^\mu} &= \sum_{\nu \neq \mu} \sum_{i=1}^{N} \underset{=0}{\underbrace{\tavg{x^\mu_i x^\nu_i}}} \tavg{\lambda^\nu} = 0, \\
    \label{eq:t_mu_variance}
    \tavg{(\eta^\mu)^2} &= \sum_{\nu,\rho \neq \mu} \sum_{i,j=1}^{N} \underset{=\delta^{\nu\rho}\delta_{ij} / N^2}{\underbrace{\tavg{x^\mu_i x^\nu_i x^\mu_j x^\rho_j}}}  \tavg{\lambda^\nu \lambda^\rho} = \alpha\tavg{(\lambda^\mu)^2}.
\end{align}
Thus, \eref{eq:local_field_split} becomes
\begin{equation}
    h^\mu = \lambda^\mu + \eta^\mu - 1, \label{eq:to_rearrange}
\end{equation}
where $\eta^\mu \sim \mathcal{N}(0, 1/\kappa^2)$ using \eref{eq:naive_capacity}. We now obtain an expression for $\lambda^\mu$ in terms of $\eta^\mu$ by applying the three as-yet unused KKT conditions. Rearranging \eref{eq:to_rearrange} and applying dual feasibility (\eref{eq:kkt_dual_feasibility}) gives $\lambda^\mu = h^\mu + 1 - \eta^\mu \geq 0$. If $\lambda^\mu > 0$, complementary slackness (\eref{eq:kkt_complementary_slackness}) requires $h^\mu = 0$, and thus $\lambda^\mu = 1 - \eta^\mu$. Alternatively, if $\lambda^\mu = 0$, from primal feasibility (\eref{eq:kkt_primal_feasibility}), we have $1 - \eta^\mu \leq 0$. Combining these conditions yields
\begin{equation}
    \label{eq:lambda_solution}
    \lambda^\mu = [1 - \eta^\mu]_+,
\end{equation}
where $[x]_+ = \text{max}(0, x)$. Inserting this expression for $\lambda^\mu$ into the capacity formula (\eref{eq:naive_capacity}) gives
\begin{equation}
    \label{eq:capacity}
    \alpha^{-1} = \tavg{\left[\kappa - z\right]_+^2}_{z \sim \mathcal{N}(0,1)},
\end{equation}
where $z = \kappa\eta^\mu$. This is the correct perceptron capacity for random points, as derived by Gardner.

While we obtain the correct final result, there is an inconsistency lurking in this naïve calculation that we now discuss.

\subsection{Inconsistency in the naïve mean-field analysis}

The naïve analysis used the relation $\tavg{w_i^2} = \alpha \tavg{(\lambda^\mu)^2}$ (\eref{eq:weight_normalization}), combined with weight-vector normalization, to derive a formula for the capacity, $\alpha^{-1} = \kappa^2\tavg{(\lambda^\mu)^2}$ (\eref{eq:naive_capacity}). However, computing the normalized L2-norm directly through the KKT conditions without disorder averaging gives 
\begin{align}
    \frac{1}{N}\sum_{i=1}^{N} w_i^2 &= \frac{1}{N}\sum_{i=1}^{N} w_i \sum_{\mu=1}^{P} \lambda^\mu x_i^\mu \label{eq:l2_start} \\
    &= \frac{1}{N} \sum_{\mu=1}^{P} (h^\mu + 1) \lambda^\mu \label{eq:l2_mid} \\
    &= \alpha \tavg{\lambda^\mu} \label{eq:l2_final},
\end{align}
where we used stationarity (\eref{eq:kkt_stationarity}) in \eref{eq:l2_start}, the local-field definition (\eref{eq:local_field}) in \eref{eq:l2_mid}, and complementary slackness (\eref{eq:kkt_complementary_slackness}) in \eref{eq:l2_final}. Weight-vector normalization therefore yields a formula for the capacity,
\begin{equation}
    \alpha^{-1} = \kappa^2 \tavg{\lambda^\mu},
    \label{eq:contradicts}
\end{equation}
contradicting \eref{eq:naive_capacity}. This inconsistency points to the fact that the naïve analysis, despite yielding the correct capacity, does not account for correlations between the dynamic variables and quenched disorder (this terminology is described below). In the next section, we describe a bipartite cavity analysis that, while only slightly more involved than the naïve method, properly handles these correlations and resolves this inconsistency.

\subsection{Bipartite cavity analysis}

The Gardner problem involves two classes of variables. The first is the \textit{quenched disorder}, given by the data $x^\mu_i$. Here, ``quenched'' means that these variables are randomly drawn, then remain fixed. The second class is \textit{dynamic variables}, given by the weights $w_i$ and Lagrange multipliers $\lambda^\mu$, which adjust to satisfy the KKT conditions. While the  naïve mean-field approach neglects the correlations between these two classes of variables, the cavity method provides a systematic way to account for their interactions. In general, the presence of correlations between the quenched disorder and dynamic variables is why techniques like the replica or cavity methods are required rather than naïve averaging-based analyses.

In addition to this general structure, the dynamic variables in the Gardner problem have a specific bipartite structure involving \textit{feature variables} $w_i$ (noting that $i=1,\ldots,N$ indexes features) and \textit{datum variables} $\lambda^\mu$ (noting that $\mu=1,\ldots,P$ indexes data) that interact through quenched random data $x_i^\mu$, which serve as couplings (\figref{fig:cavity_fig}, ``Gardner capacity''). We describe this bipartite interaction using the equations
\begin{align}
    h^\mu &= \sum_{i=1}^{N} x_i^\mu w_i -1 + I^\mu, \label{eq:h_mu} \\
    w_i &= \sum_{\mu=1}^{P} x_i^\mu \lambda^\mu + I_i \label{eq:w_i},
\end{align}
where $I^\mu$ and $I_i$ are infinitesimal source terms that we have added for computing response functions. The system description is completed by the remaining three KKT conditions (Eqs.~\ref{eq:kkt_primal_feasibility}--\ref{eq:kkt_complementary_slackness}). When the source terms are zero, the quenched disorder, $x^\mu_i$, uniquely determines the dynamic variables, $w_i$ and $\lambda^\mu$, assuming that the subset of points $x^\mu_i$ that are supporting, i.e., have $\lambda^\mu > 0$, are in general position. For random points, the probability of these supporting points not being in general position is zero. We assume that this uniqueness property holds in the presence of the infinitesimal source terms.

The bipartite cavity method involves performing the following steps for both feature and datum variables:
\begin{enumerate}
    \item Start with an unperturbed system: dynamic variables, $w_i$ and $\lambda^\mu$, for a given realization of quenched disorder, $x^\mu_i$.
    \item Introduce a new ``cavity'' dynamic variable that perturbs the existing dynamic variables. For features, this is a new weight $w_0$; for data, a new Lagrange multiplier $\lambda^0$. This requires introducing new quenched random variables, $x^\mu_0$ or $x^0_i$, respectively, that couple the cavity variable to the existing variables.
    \item Write an equation describing how this cavity variable responds to the other variables, accounting for how those variables were perturbed by the cavity variable's introduction.
    \item Solve the resulting cavity equations self-consistently by averaging over the quenched disorder.
\end{enumerate}
The bipartite structure simplifies this analysis: when introducing a cavity dynamic variable (feature or datum), we only need to compute its effect on the opposite set of dynamic variables (datum or feature, respectively). This is because only the opposite set provides input to the cavity variable in step 3.

We repeat these steps separately for both the feature-cavity and data-cavity cases. This produces two complementary cavity pictures, one for each set of dynamic variables. The self-consistent equations in each picture depend on statistical averages in the other picture.

We now implement these steps for the Gardner problem.

\subsubsection{Feature cavity}

We begin with unperturbed dynamic variables $w_i$ and $\lambda^\mu$, then introduce a new weight $w_0$, with quenched random variables $x_0^\mu$ that connect it to the existing Lagrange multipliers $\lambda^\mu$. Since the couplings are weak ($x^\mu_i \sim 1/\sqrt{N}$), introducing $w_0$ delivers small perturbations to the existing Lagrange multipliers, given by
\begin{equation}
    \label{eq:delta_lambda_mu}
    \delta \lambda^\mu = \sum_{\nu=1}^P \frac{d \lambda^\mu}{d I^\nu} x_0^\nu w_0.
\end{equation}
One could include higher-order terms, the next of which would be smaller than the above term by a factor of $1/\sqrt{N}$ and given by $\sum_{\nu,\rho=1}^P \frac{d\lambda^\mu}{dI^\nu dI^\rho} x^\nu_0 x^\rho_0 w_0^2$. Tracking such higher-order terms is typically not necessary. 

Upon introduction of $w_0$, there is also a perturbation to $w_i$ for $i = 1,\ldots,N$, however these perturbations do not enter directly into the expression for $w_0$ due to the bipartite structure of the interactions. In particular, given the perturbation \eref{eq:delta_lambda_mu} above, $w_0$ follows
\begin{align}
    w_0 &= \sum_{\mu=1}^{P} x_0^\mu (\lambda^\mu + \delta \lambda^\mu) + I_0 \label{eq:w0_start} \\
    &= \sum_{\mu=1}^{P} x_0^\mu \lambda^\mu + \sum_{\mu,\nu=1}^{P} x_0^\mu x_0^\nu  \frac{d \lambda^\mu}{d I^\nu} w_0+ I_0 \label{eq:w0_mid} \\
    &= \eta_0 + F_{00} w_0+ I_0 \label{eq:w0_dynamics},
\end{align}
where we define the cavity field $\eta_0$ and self-coupling $F_{00}$ as
\begin{align}
    \eta_0 &= \sum_{\mu=1}^{P} x_0^\mu \lambda^\mu, \label{eq:eta0} \\
    F_{00} &= \sum_{\mu,\nu=1}^{P} x_0^\mu  x_0^\nu \frac{d \lambda^\mu}{d I^\nu} \label{eq:F00}.
\end{align}
The cavity construction ensures that, in these expressions, \textit{the quenched disorder, $x_0^\mu$, and dynamic variables, $\lambda^\mu$ and ${d \lambda^\mu}/{d I^\nu}$, are independent of each other}. This solves the issue in the naïve analysis and allows for the calculation of disorder-averaged statistics of these expressions. Applying the central limit theorem, $\eta_0$ is Gaussian with
\begin{align}
    \tavg{\eta_0} &= 0,  \\
    \tavg{\eta_0^2} &= \alpha \tavg{(\lambda^\mu)^2}.
\end{align}
Meanwhile, the self-coupling $F_{00}$ is order-one and self-averaging with mean
\begin{equation}
    \tavg{F_{00}} = \alpha S^{\lambda} , \label{eq:F00exp}
\end{equation}
where $S^\lambda = \tavg{{d \lambda^\mu}/{d I^\mu}}$ is the average on-diagonal Lagrange-multiplier response. By self-averaging, we mean that the fluctuations of $F_{00}$ (defined by \eref{eq:F00}) around its $\mathcal{O}(1)$ expectation (given by \eref{eq:F00exp}) are $\mathcal{O}(1/\sqrt{N})$, as can be verified. Next, from \eref{eq:w0_dynamics}, we obtain
\begin{equation}
    w_0 = \eta_0 + \alpha S^\lambda w_0 + I_0 \label{eq:w0_solution},
\end{equation}
leading to the weight response function
\begin{equation}
    S^{w} = \frac{d w_0}{d I_0 } = \frac{1}{1 - \alpha S^{\lambda}} \label{eq:S_w}.
\end{equation}
Dropping the source term $I_0$, we have
\begin{equation}
    w_0 = S^w \eta_0 \label{eq:w0_final}.
\end{equation}
Thus, weight-vector normalization $\tavg{w_0^2} = 1/\kappa^2$ yields the capacity formula
\begin{equation}
    \alpha^{-1} = \kappa^2 (S^w)^2  \tavg{(\lambda^\mu)^2} \label{eq:normalization}.
\end{equation}
To complete the calculation, we perform the same analysis with a datum cavity variable $\lambda^0$. This will allow us to determine $\tavg{(\lambda^\mu)^2} = \tavg{(\lambda^0)^2}$.


\subsubsection{Datum cavity}

We begin with unperturbed dynamic variables $w_i$ and $\lambda^\mu$, then introduce a new Lagrange multiplier $\lambda^0$ with quenched random variables $x_i^0$ that connect it to the existing weights $w_i$. As before, since the couplings are weak, introducing $\lambda^0$ delivers small perturbations to the existing weights,
\begin{equation}
    \label{eq:delta_w_i}
    \delta w_i = \sum_{j=1}^N \frac{d w_i}{d I_j} x_j^0 \lambda^0.
\end{equation}
The bipartite structure implies that the perturbations to other Lagrange multipliers $\lambda^\mu$ with $\mu = 1,\ldots, P$ do not enter directly into the expression for $h^0$, which reads
\begin{align}
    h^0 &= \sum_{i=1}^{N} x_i^0 (w_i + \delta w_i) - 1 + I^0 \label{eq:h0_start} \\
    &= \sum_{i=1}^{N} x_i^0 w_i + \sum_{i,j=1}^{N} x_i^0 x_j^0 \frac{d w_i}{d I_j} \lambda^0 - 1 + I^0 \label{eq:h0_mid} \\
    &= \eta^0 + F^{00} \lambda^0 - 1 + I^0 \label{eq:h0_dynamics},
\end{align}
where we define the cavity field $\eta^0$ and self-coupling $F^{00}$ as
\begin{align}
    \eta^0 &= \sum_{i=1}^{N} x_i^0 w_i \label{eq:eta0_data}, \\
    F^{00} &= \sum_{i,j=1}^{N} x_i^0 x_j^0 \frac{d w_i}{d I_j} \label{eq:F00_data}.
\end{align}
The cavity construction ensures that, in these expressions, \textit{the quenched disorder, $x_i^0$, and dynamic variables, $w_i$ and ${d w_i}/{d I_j}$, are independent of each other}, allowing for calculation of disorder-averaged statistics of these expressions. Applying the central limit theorem,  $\eta^0$ is Gaussian with
\begin{align}
    \tavg{\eta^0} &= 0 \label{eq:eta0_data_mean}, \\
    \tavg{(\eta^0)^2} &= \tavg{w_i^2} = \frac{1}{\kappa^2} \label{eq:eta0_data_var}.
\end{align}
Meanwhile, the self-coupling $F^{00}$ is order-one and self-averaging with mean
\begin{equation}
    \tavg{F^{00}} = S^w \label{eq:F00_data_mean}.
\end{equation}
From \eref{eq:h0_dynamics}, we obtain
\begin{equation}
    h^0 = \eta^0 + S^w \lambda^0 - 1 + I^0 \label{eq:h0_solution}.
\end{equation}
Using the same logic based on the KKT conditions as in the naïve analysis, the cavity Lagrange multiplier follows
\begin{equation}
    \lambda^0 = \frac{[1 - \eta^0 - I^0]_+}{S^w} \label{eq:lambda0_final}.
\end{equation}
Dropping the source term $I^0$ and inserting this expression into the capacity formula (\eref{eq:normalization}), we obtain
\begin{equation}
    \alpha^{-1} = \kappa^2 (S^w)^2 \tavg{\frac{[1 - \eta^0]_+^2}{(S^w)^2}} \label{eq:normalization_gardner}.
\end{equation}
The factors of $(S^w)^2$ cancel, yielding the Gardner result (\eref{eq:capacity}).

\subsection{Symmetry in the problem}
\label{subsec:symmetry}

Clearly, the success of the naïve mean-field analysis relied on the fortuitous cancellation of the factors $(S^w)^2$. Why did this happen?

Consider generalizing the weight equation (\eref{eq:w_i}) to
\begin{equation}
    w_i = A\left(\sum_{\mu=1}^{P} x_i^\mu \lambda^\mu + I_i\right),
\end{equation}
where $A > 0$ is a scalar. For any max-margin solution with support-vector coefficients $\lambda^\mu$ at $A=1$, there exists a solution with $\lambda^{\mu'} = \lambda^\mu / A$ for any positive $A$. Thus $A$ does not affect capacity, though it scales the linear response function $S^w$. The capacity result therefore cannot depend on $S^w$, and one is free to assume for it any positive value, for instance $S^w = 1$. This symmetry in the bipartite description of the problem explains why the naïve mean-field calculation succeeded: it implicitly assumed $S^w = 1$ when calculating both $\tavg{w_i^2}$ and $h^\mu$. 

\subsection{Computing the response functions}
While the capacity calculation is independent of $S^w$, this response function is required to determine other important quantities, such as the distribution of Lagrange multipliers. Moreover, computing $S^w$ allows us to verify that the cavity method resolves the inconsistency found in the naïve analysis.

From the cavity solution for $\lambda^0$ (\eref{eq:lambda0_final}), we obtain the Lagrange-multiplier response
\begin{equation}
    S^\lambda = -\frac{\phi_0(\kappa)}{S^w} \label{eq:S_lambda_phi},
\end{equation}
where we define the function
\begin{equation}
    \phi_n(\kappa) = \int_{-\infty}^\kappa \mathcal{D}z (\kappa - z)^n,
\end{equation}
with $\mathcal{D}z = (2\pi)^{-1/2}\exp(-z^2/2)$ denoting Gaussian measure. Combining this with the weight response (\eref{eq:S_w}) yields
\begin{align}
    S^\lambda &= -\frac{\phi_0(\kappa)}{1 - \alpha \phi_0(\kappa)} \label{eq:S_lambda_sol}, \\
    S^w &= 1 - \alpha \phi_0(\kappa) \label{eq:S_w_sol}.
\end{align}
Consulting \eref{eq:lambda0_final}, $\phi_0(\kappa)$ is the probability for a Lagrange multiplier to be nonzero. Thus, the weight response $S^w$ (\eref{eq:S_w_sol}) has a nice interpretation:
\begin{equation}
    S^w = 1 - \frac{\text{number of supporting points}}{N}.
\end{equation}
This implies that the response to adding a new datum is strongest when there are few supporting points. This makes intuitive sense: with fewer supporting points, each point has a greater influence on the weight vector. As the number of supporting points increases, the influence of each individual point diminishes. 

\subsection{Resolving the inconsistency}
Having derived the response functions, we now show that the inconsistency in the naïve calculation is resolved. Recall the two formulas for the capacity:
\begin{equation}
    \alpha^{-1} = \kappa^2 (S^w)^2 \tavg{(\lambda^\mu)^2} = \kappa^2 \tavg{\lambda^\mu}, \label{eq:apparent_contradiction}
\end{equation}
from Eqs.~\ref{eq:weight_normalization} and \ref{eq:contradicts}. These expressions appeared contradictory only when we naïvely and incorrectly assumed $S^w = 1$. We now show that both expressions are consistent when using the correct value of $S^w = 1- \alpha \phi_0(\kappa)$ derived from the cavity analysis.

Substituting the expression for $\lambda^0$ from \eref{eq:lambda0_final} into \eref{eq:apparent_contradiction}, we obtain
\begin{equation}
    \alpha^{-1} = \phi_2(\kappa) = \left[1 - \frac{\phi_0(\kappa)}{\phi_2(\kappa)}\right]^{-1} \kappa \phi_1(\kappa),
\end{equation}
which can be rearranged to yield the identity
\begin{equation}
    \phi_2(\kappa) = \phi_0(\kappa) + \kappa \phi_1(\kappa). \label{eq:relation}
\end{equation}
We can verify this identity through direct calculation: the left-hand side can be written as
\begin{align}
    \phi_2(\kappa) &= - \int_{-\infty}^{\kappa} \mathcal{D}z z (\kappa - z) + \kappa \phi_1(\kappa) \\
    &= \int_{-\infty}^{\kappa} \mathcal{D}z + \kappa \phi_1(\kappa) \\
    &= \phi_0(\kappa) + \kappa \phi_1(\kappa),
\end{align}
where we used integration by parts in the second step.

\section{Manifold capacity}

\subsection{Background}
The Gardner problem studies the separability of random points, but real-world data typically exhibit complex geometric structure. For example, images of an object captured from different angles form a continuous manifold in pixel space. More generally, inputs that correspond to the same output (forming a ``class'') create structured manifolds, either in the input space or in some transformed representation. Certain neural circuits (e.g., sensory systems) and artificial neural systems (e.g., deep-network classifiers) must generate outputs that remain invariant to such input transformations. The manifold capacity problem extends Gardner's analysis from discrete points to continuous manifolds, addressing this reality. Chung et al. \cite{chung2018classification} first solved this manifold capacity problem, building on their earlier work on specific manifold geometries \cite{chung2016linear}. While their paper outlined the capacity calculation using a na\"ive mean-field approach, we present both a detailed naïve analysis and a complete bipartite cavity analysis. As in the Gardner case, the naïve calculation yields the correct capacity due to the presence of a symmetry in the problem description.

\subsection{Problem formulation}
Consider a dataset of $P$ manifolds, each with a base shape $S \subseteq \mathbb{R}^{D+1}$ and binary label $y^\mu \in \{-1, +1\}$. Each manifold is embedded in $N$-dimensional space through random isometric embeddings indexed by $\mu = 1, \ldots, P$. The embedding matrices have components $u^\mu_{ia}$, where $i=1,\ldots,N$ indexes features and $a = 1,\ldots,D+1$ indexes the manifold subspace dimension. We denote the $i$-th row of this embedding matrix by $\bm{u}^\mu_i$, reserving bold notation for $(D+1)$-dimensional vectors and matrices. A point with intrinsic coordinates $\bm{s} \in S$ on manifold $\mu$ has $N$-dimensional coordinates given by
\begin{equation}
    \label{eq:manifold_point}
    x_i^\mu(\bm{s}) = \sum_{a=1}^{D+1} u^{\mu}_{ia} s_a = (\bm{u}_i^\mu)^T \bm{s}.
\end{equation}
We assume that the origin is not in the convex hull of $S$, ensuring that linear separability of the manifolds is possible. A convenient way to parameterize the base shape, though not required for the present analysis, is to use the last coordinate $s_{D+1}$ to specify the displacement from the origin, with the remaining coordinates $(s_1, \ldots, s_{D})$ specifying the shape of the manifold in a $D$-dimensional subspace orthogonal to the displacement (e.g., a sphere with radius $R$ a distance $d$ from the origin would be specified by $s_{D+1} = d$, $s_1^2 + \ldots + s_D^2 = R^2$). Since we will consider linear classification of manifolds, the problem is only sensitive to the convex hull of each manifold, and thus $S$ can be replaced with the convex hull of $S$ without loss of generality. 

The classification task requires finding a weight vector with components $w_i$ that separates manifolds into their assigned classes. Specifically, every point $x^\mu_i(\bm{s})$ on manifold $\mu$ must satisfy
\begin{equation}
    \label{eq:manifold_classification}
    y^\mu \sum_{i=1}^{N} w_i x_i^\mu(\bm{s}) \geq 1.
\end{equation}
To analyze this condition more compactly, we introduce the vectors
\begin{equation}
    \label{eq:V_mu_def}
    \bm{V}^\mu = y^\mu \sum_{i=1}^{N} w_i \bm{u}_i^\mu.
\end{equation}
This allows us to express the classification condition as $(\bm{V}^\mu)^T \bm{s} \geq 1$ for all $\bm{s} \in S$. We can further simplify this using the \textit{support function} of $S$, a fundamental concept in convex geometry\footnote{This ``min'' definition of the support function is trivially related to the more conventional ``max'' definition by negating its argument \cite{chung2018classification}.}:
\begin{equation}
    \label{eq:support_function}
    g_S(\bm{y}) = \min_{\bm{s} \in S} \bm{y}^T \bm{s}.
\end{equation}
This leads to a manifold analogue of the local field,
\begin{equation}
    \label{eq:manifold_field}
    h^\mu = g_S(\bm{V}^\mu) - 1.
\end{equation}
The classification condition then becomes simply $h^\mu \geq 0$, as in the Gardner case.

As before, we formulate an optimization problem that minimizes the weight-vector norm subject to correct classification. The Lagrangian takes the same form as \eref{eq:lagrangian}, with Lagrange multipliers $\lambda^\mu$ multiplying $h^\mu$. To write down the KKT conditions, we introduce the concept of an \textit{anchor point} for each manifold, defined by the gradient of the support function:
\begin{equation}
    \label{eq:anchor_point}
    \bm{s}^\mu = \nabla g_S(\bm{V}^\mu).
\end{equation}
A key fact from convex geometry is that the gradient of this support function gives the ``arg-min,'' over the manifold, of the inner product with $\bm{V}^\mu$,
\begin{equation}
    \label{eq:anchor_point_alt}
    \bm{s}^\mu = \underset{\bm{s} \in S}{\text{arg-min}} (\bm{V}^\mu)^T \bm{s}.
\end{equation}
Thus, the anchor point $s^\mu$ is the intrinsic coordinate of the point on manifold $\mu$ that is closest to the decision boundary. With this definition, the KKT stationarity condition ($d\mathcal{L}/d w_i = 0$) reads
\begin{align}
    w_i &= \sum_{\mu=1}^{P} \lambda^\mu y^\mu (\bm{u}_i^\mu)^T\bm{s}^\mu, \label{eq:manifold_kkt_stationarity}
\end{align}
with the remaining conditions (primal feasibility, dual feasibility, and complementary slackness) identical to Eqs.~\ref{eq:kkt_primal_feasibility}--\ref{eq:kkt_complementary_slackness} from the Gardner case.

\subsection{Random-embedding assumption}
In place of random points, the manifold capacity problem assumes random embeddings of the manifolds. These embedding components $u^\mu_{ia}$ are i.i.d. with
\begin{align}
    \label{eq:embedding_mean}
    \tavg{u^\mu_{ia}} &= 0, \\
    \label{eq:embedding_var}
    \tavg{u^\mu_{ia} u^\nu_{jb} } &= \frac{1}{N} \delta^{\mu\nu} \delta_{ij} \delta_{ab}.
\end{align}
This scaling ensures that each manifold's embedding is isometric for $N \to \infty$ and finite $D$. As in the Gardner case, we set $y^\mu = 1$ for all $\mu$ without loss of generality. The precise form of $P(u^\mu_{ia})$ does not enter into the calculation due to the central limit theorem.

\subsection{Naïve mean-field analysis}
In analogy with the Gardner case, we begin with a concise naïve mean-field analysis. While this analysis makes incorrect independence assumptions, it yields the correct capacity due to the same underlying symmetry present in the Gardner case.

Starting from the stationarity condition (\eref{eq:manifold_kkt_stationarity}), we incorrectly assume independence between $\lambda^\mu \bm{s}^\mu$ and $\bm{u}^\mu_i$, yielding
\begin{align}
    \tavg{w_i^2} &= \sum_{\mu,\nu=1}^{P} \sum_{a,b=1}^{D+1} \underset{=\delta_{ab}\delta^{\mu\nu}/N}{\underbrace{\tavg{u^\mu_{ia} u^\nu_{ib}}}} \tavg{\lambda^\mu s^\mu_a \lambda^\nu s^\nu_b} \\
    &= \alpha \tavg{ \lVert \lambda^\mu \bm{s}^\mu \rVert_2^2}. \label{eq:manifold_norm_thing}
\end{align}
From weight-vector normalization $\tavg{w_i^2} = 1/\kappa^2$, we obtain a formula for the capacity,
\begin{equation}
    \label{eq:naive_manifold_capacity}
    \alpha^{-1} = \kappa^2\tavg{ \lVert\lambda^\mu \bm{s}^\mu \rVert_2^2}.
\end{equation}
Next, we analyze the vector $\bm{V}^\mu$ by substituting \eref{eq:manifold_kkt_stationarity} into \eref{eq:V_mu_def}, yielding
\begin{equation}
    \bm{V}^\mu = \sum_{\nu=1}^P  \sum_{i=1}^{N} \bm{u}^\mu_i (\bm{u}^{\nu}_i)^T \lambda^\nu \bm{s}^{\nu}.
\end{equation}
Separating diagonal ($\mu=\nu$) and off-diagonal ($\mu\neq\nu$) terms gives
\begin{equation}
    \label{eq:V_split}
    \bm{V}^\mu = {\left[ \sum_{i=1}^N \bm{u}^\mu_i (\bm{u}^\mu_i)^T \right]} \lambda^\mu \bm{s}^\mu + {\sum_{\nu\neq\mu} \sum_{i=1}^N \bm{u}^\mu_i (\bm{u}^\nu_i)^T \lambda^\nu \bm{s}^\nu}.
\end{equation}
The square-bracket term is $\bm{\mathcal{I}}$ (the identity matrix) at large $N$ due to the choice of embedding variance (\eref{eq:embedding_var}). Applying again the incorrect assumption of independence between $x^\mu_i$ and $\lambda^\mu$, the second term, which we denote by $\bm{T}^\mu = {\sum_{\nu\neq\mu} \sum_{i=1}^N \bm{u}^\mu_i (\bm{u}^\nu_i)^T \lambda^\nu \bm{s}^\nu}$, is Gaussian (by the central limit theorem) with
\begin{align}
    \label{eq:T_mean}
    \tavg{{T}^\mu_a} &= \sum_{\nu\neq\mu} \sum_{i=1}^{N} \sum_{b=1}^{D+1} \underset{=0}{\underbrace{\tavg{u^\mu_{ia} u^\nu_{ib}}}} \tavg{\lambda^\nu s^\nu_b} = 0, \\ 
    \label{eq:T_var}
    \tavg{{T}^\mu_a {T}^\mu_b} &= \sum_{\nu,\rho \neq \mu}\sum_{i,j=1}^{N} \sum_{c,d=1}^{D+1} \underset{=\delta^{\nu\rho}\delta_{ij}\delta_{ab} \delta_{cd} /N^2}{\underbrace{\tavg{u^\mu_{ia} u^\nu_{ic} u^\mu_{jb} u^\rho_{jd}}}} \tavg{ \lambda^\nu s^\nu_c \lambda^\rho s^\rho_d} \\
    &= \alpha \tavg{\lVert \lambda^\mu \bm{s}^\mu \rVert_2^2} \delta_{ab}.
\end{align}
Thus, from \eref{eq:V_split}, we have
\begin{equation}
    \label{eq:V_final}
    \bm{V}^\mu = \bm{T}^\mu + \lambda^\mu \bm{s}^\mu,
\end{equation}
where $\bm{T}^\mu \sim \mathcal{N}(\bm{0}, \bm{\mathcal{I}}/\kappa^2)$ (using \eref{eq:manifold_norm_thing}). This leads to the capacity result
\begin{equation}
    \label{eq:final_manifold_capacity}
    \alpha^{-1} = \kappa^2\tavg{\lVert \bm{V}^\mu - \bm{T}^\mu \rVert_2^2}_{\bm{T}^\mu \sim \mathcal{N}(\bm{0}, \bm{\mathcal{I}}/\kappa^2)}.
\end{equation}
Given a sample of $\bm{T}^\mu$, the corresponding $\bm{V}^\mu$ is determined by the KKT conditions, in analogy to how $\lambda^\mu$ was determined from $\eta^\mu$ in the Gardner case. Specifically, \eref{eq:V_final}, together with the primal feasibility, dual feasibility, and complementary slackness conditions, comprise the KKT conditions for the convex optimization problem
\begin{equation}
\label{eq:manifold_opt_final}
\bm{V}^\mu = \underset{\bm{V}}{\text{arg min}} \frac{1}{2} \lVert \bm{V} - \bm{T}^\mu \rVert_2^2 \quad \text{subject to } g_S(\bm{V}) \geq 1,
\end{equation}
which has a unique solution $\bm{V}^\mu$ for differentiable manifolds $S$. \eref{eq:final_manifold_capacity} is the correct capacity for random manifolds derived by Chung et al. \cite{chung2018classification}.

As in the Gardner problem, there is an inconsistency lurking in this naïve mean-field analysis that results in two contradictory formulas for the capacity (the other being $\alpha^{-1} = \kappa^2 \tavg{\lambda^\mu}$). In the Appendix, we show that this is resolved through the bipartite cavity analysis presented next, which properly handles the correlations between variables and allows for calculation of response functions. 

\subsection{Bipartite cavity analysis}
The cavity method properly handles correlations between the quenched disorder and dynamic variables. We analyze a bipartite system where the interactions are mediated through random embeddings (\figref{fig:cavity_fig}, ``Manifold capacity''). The system is described by
\begin{align}
    \label{eq:manifold_V_dynamics}
    \bm{V}^\mu &= \sum_{i=1}^{N} w_i \bm{u}_i^\mu + \bm{I}^\mu, \\
    \label{eq:manifold_w_dynamics}
    w_i &= \sum_{\mu=1}^{P} (\bm{u}_i^\mu)^T(\lambda^\mu \bm{s}^\mu) + I_i,
\end{align}
where $\bm{I}^\mu$ and $I_i$ are infinitesimal source terms for computing response functions. The feature variables are weights $w_i$, as in the Gardner case, while the datum variables are now vector-valued products $\lambda^\mu \bm{s}^\mu$, which can be interpreted as vector generalizations of the Lagrange multipliers from the Gardner case. We call them vector-Lagrange multipliers for this reason. 

The cavity analysis proceeds in parallel for features and data, following the same four steps as before: starting with unperturbed variables, introducing a cavity variable, analyzing its response to existing variables, and solving self-consistently through disorder averaging. We now do each case in turn.

\subsubsection{Feature cavity}

We begin with unperturbed dynamic variables $w_i$ and $\lambda^\mu\bm{s^\mu}$, then introduce a new weight $w_0$ with quenched random variables $\bm{u}_0^\mu$ that connect it to the existing vector-Lagrange multipliers $\lambda^\mu\bm{s}^\mu$. Since the couplings are weak ($u^\mu_{ia} \sim 1/\sqrt{N}$), introducing $w_0$ delivers small perturbations to the existing vector-Lagrange multipliers,
\begin{equation}
    \label{eq:manifold_delta_lambda}
    \delta(\lambda^\mu \bm{s}^\mu) = \sum_{\nu=1}^{P} \frac{d(\lambda^\mu \bm{s}^\mu)}{d\bm{I}^\nu} w_0 \bm{u}_0^\nu,
\end{equation}
where $[d(\lambda^\mu \bm{s}^\mu)/d\bm{I}^\nu]_{ab} = d(\lambda^\mu s^\mu_a)/dI^\nu_b$ is a response matrix. The bipartite structure implies that the perturbations to other weights $w_i$ with $i = 1, \ldots, N$ do not enter directly into the expression for $w_0$, which reads
\begin{align}
    w_0 &= \sum_{\mu=1}^{P} (\bm{u}_0^\mu)^T[\lambda^\mu \bm{s}^\mu + \delta(\lambda^\mu \bm{s}^\mu)] + I_0  \\
    &= \sum_{\mu=1}^{P} (\bm{u}_0^\mu)^T \lambda^\mu \bm{s}^\mu + \sum_{\mu,\nu=1}^{P} (\bm{u}_0^\mu)^T \frac{d(\lambda^\mu \bm{s}^\mu)}{d\bm{I}^\nu} \bm{u}_0^\nu w_0 + I_0  \\
    &= \eta_0 + F_{00}w_0 + I_0, \label{eq:manifold_w0_dynamics}
\end{align}
where we define the cavity field $\eta_0$ and self-coupling $F_{00}$ as
\begin{align}
    \label{eq:manifold_eta0}
    \eta_0 &= \sum_{\mu=1}^{P} (\bm{u}_0^\mu)^T \lambda^\mu \bm{s}^\mu, \\
    \label{eq:manifold_F00}
    F_{00} &= \sum_{\mu,\nu=1}^{P} (\bm{u}_0^\mu)^T \frac{d(\lambda^\mu \bm{s}^\mu)}{d\bm{I}^\nu} \bm{u}_0^\nu.
\end{align}
The cavity construction ensures that, in these expressions, \textit{the quenched disorder, $\bm{u}_0^\mu$, and dynamic variables, $\lambda^\mu \bm{s}^\mu$ and ${d(\lambda^\mu \bm{s}^\mu)}/{d\bm{I}^\nu}$, are independent of each other}, allowing for calculation of disorder-averaged statistics of these expressions. Applying the central limit theorem, $\eta_0$ is Gaussian with
\begin{align}
    \label{eq:manifold_eta0_stats}
    \tavg{\eta_0} &= 0, \\
    \tavg{\eta_0^2} &= \alpha \tavg{ \lVert \lambda^\mu\bm{s}^\mu \rVert_2^2}.
\end{align}
Meanwhile, the self-coupling is order-one and self-averaging with mean
\begin{equation}
    \label{eq:manifold_F00_mean_manifold}
    \tavg{F_{00}} = \alpha S^\lambda,
\end{equation}
where we define
\begin{equation}
    S^\lambda = \tavg{\text{tr}\frac{d(\lambda^\mu \bm{s}^\mu)}{d\bm{I}^\mu}}. 
\end{equation}
From \eref{eq:manifold_w0_dynamics}, we obtain $w_0 = \eta_0 + \alpha S^\lambda w_0 + I_0$ and thus the weight response function $ S^w = 1/({1 - \alpha S^\lambda})$, identical to the Gardner case (\eref{eq:S_w}). Dropping source terms, we have $w_0 = S^w \eta_0$. Weight-vector normalization $\tavg{w_0^2} = 1/\kappa^2$ then yields the capacity formula
\begin{equation}
    \label{eq:manifold_capacity_condition}
    \alpha^{-1} = \kappa^2 (S^w)^2\tavg{\lVert \lambda^\mu\bm{s}^\mu \rVert_2^2}.
\end{equation}
We now finish the calculation by performing a datum cavity analysis. 

\subsubsection{Datum cavity}

For the datum cavity analysis, we begin with unperturbed dynamic variables $w_i$ and $\lambda^\mu$, then introduce a new vector-Lagrange multiplier $\lambda^0\bm{s}^0$ with quenched random variables $\bm{u}_i^0$ that connect it to the existing weights $w_i$. Since the couplings are weak, introducing $\lambda^0 \bm{s}^0$ delivers small perturbations to the existing weights,
\begin{equation}
    \label{eq:manifold_delta_w}
    \delta w_i = \sum_{j=1}^N \frac{dw_i}{dI_j} (\bm{u}_j^0)^T(\lambda^0 \bm{s}^0).
\end{equation}
The bipartite structure implies that the perturbations to other vector-Lagrange multipliers $\lambda^\mu \bm{s}^\mu$ with $\mu = 1, \ldots, P$ do not enter directly into the expression for $\bm{V}^0$, which reads
\begin{align}
    \bm{V}^0 &= \sum_{i=1}^{N} (w_i + \delta w_i )\bm{u}^0_i + \bm{I}^0  \\
    &= \sum_{i=1}^{N} w_i \bm{u}^0_i + \sum_{i,j=1}^{N} \frac{dw_i}{dI_j} \bm{u}^0_i (\bm{u}_j^0)^T(\lambda^0 \bm{s}^0) + \bm{I}^0  \\
    &= \bm{T}^0 + \bm{F}^{00} \lambda^0 \bm{s}^0 + \bm{I}^0, \label{eq:manifold_V0_dynamics}
\end{align}
where we define the cavity field $\bm{T}^0$ and self-coupling matrix $\bm{F}^{00}$ as
\begin{align}
    \label{eq:manifold_T0}
    \bm{T}^0 &= \sum_{i=1}^{N} w_i \bm{u}^0_i, \\
    \label{eq:manifold_F00_matrix}
    \bm{F}^{00} &= \sum_{i,j=1}^{N} \frac{dw_i}{dI_j} \bm{u}^0_i (\bm{u}_j^0)^T.
\end{align}
The cavity construction ensures that, in these expressions, \textit{the quenched disorder, $\bm{u}^0_i$, and dynamic variables, $w_i$ and ${dw_i}/{dI_j}$, are independent of each other}, allowing for calculation of disorder-averaged statistics of these expressions. Applying the central limit theorem,  $\bm{T}^0$ is Gaussian with
\begin{align}
    \label{eq:manifold_T0_stats}
    \tavg{{T}^0_a} &= 0, \\
    \tavg{{T}^0_a {T}^0_b} &= \frac{1}{\kappa^2} \delta_{ab},
\end{align}
The self-coupling matrix is order-one and self-averaging with mean
\begin{equation}
    \label{eq:manifold_F00_matrix_mean}
    \tavg{{F}^{00}_{ab}} = S^w \delta_{ab}.
\end{equation}
From \eref{eq:manifold_V0_dynamics}, we obtain
\begin{equation}
    \label{eq:manifold_V0_solution}
    \bm{V}^0 = \bm{T}^0 + S^w \lambda^0 \bm{s}^0 + \bm{I}^0.
\end{equation}
Using the same logic regarding the KKT conditions as in the naïve analysis, we recognize that \eref{eq:manifold_V0_solution}, together with the primal feasibility, dual feasibility, and complementary slackness conditions comprise the KKT conditions for the convex optimization problem
\begin{equation}
    \label{eq:manifold_opt_final_cavity}
    \bm{V}^{0} = \underset{\bm{V}}{\text{arg min}} \: \frac{1}{2} \lVert \bm{V} - \bm{T}^{0} \rVert_2^2 \quad \text{subject to } g_S(\bm{V}) \geq 1,
\end{equation}
which we have reproduced from the naïve analysis (\eref{eq:manifold_opt_final}) with the cavity notation, and which has a unique solution for differentiable manifolds. Upon insertion into the capacity formula (\eref{eq:manifold_capacity_condition}), we obtain
\begin{equation}
    \label{eq:manifold_final_capacity}
    \alpha^{-1} = \kappa^2 (S^w)^2 \tavg{\frac{\lVert \bm{V}^{0} - \bm{T}^{0} \rVert_2^2}{(S^w)^2}}.
\end{equation}
The $(S^w)^2$ factors cancel, yielding the correct capacity of \eref{eq:final_manifold_capacity}. As in the Gardner problem, this cancellation reflects an underlying symmetry in the bipartite description of the problem.

The complete calculation of response functions and resolution of the inconsistency between capacity formulas in the naïve analysis is given in the Appendix.

\section{Kernel ridge regression}

\subsection{Background}

Having analyzed two classification problems, we now turn to regression, where the goal is to predict continuous outputs rather than binary labels. In particular, we consider kernel regression, which extends linear regression to nonlinear function approximation while preserving the mathematical tractability of linear methods. The key insight is to implicitly map input data into a high-dimensional (often, infinite-dimensional) feature space through a kernel function that computes inner products between these implicit feature vectors without explicitly constructing them. This ``kernel trick'' allows the method to learn nonlinear relationships in the original input space. A ridge penalty controls the complexity of the learned function by regularizing its norm, promoting generalization to new data.

Despite having an analytical solution, understanding how kernel ridge regression generalizes---that is, how well it performs on unseen data as a function of training set size, data properties, and regularization strength---is nontrivial. The generalization error was first computed using the replica method by Bordelon, Canatar, and Pehlevan \cite{bordelon2020spectrum, canatar2021spectral} (see also \cite{spigler2020asymptotic}, which derived similar results in a specific setting; as well as earlier analyses of Gaussian process regression \cite{sollich1998learning, sollich2002learning}). Simon et al. \cite{simon2023eigenlearning} later derived equivalent results through a conservation law, using a cavity-like argument based on adding a single kernel eigenmode. The cavity analysis presented here differs by introducing a bipartite structure that considers both the addition of a kernel eigenmode (feature cavity) and a training example (datum cavity). This approach parallels recent work by Bordelon et al. \cite{bordelon2024dynamical}, who used a cavity method, in addition to a functional-integral method, to derive dynamical mean-field equations for learning dynamics, with the steady state describing the generalization error calculated here (see also \cite{agoritsas2018out}). Other work has derived this result using random matrix theory \cite{jacot2020kernel}. Atanasov et al. \cite{atanasov2024scaling} recently reviewed these results and connected them to renormalization concepts from physics. The insights from kernel ridge regression illuminate generalization in deep-learning systems through frameworks like the neural tangent kernel \cite{jacot2018neural}. In contrast to the typical-case behavior, \cite{caponnetto2007optimal} considered worst-case behavior. Cui et al. \cite{cui2021generalization} showed that differences in kernel ridge regression decay rates previously attributed to typical-case vs. worst-case analyses actually result from noiseless vs. noisy data assumptions. 

\subsection{Problem formulation}

Consider a supervised learning task with $P$ training examples $x^\mu \in \chi$ from an arbitrary input space $\chi$ and corresponding real-valued targets $y^\mu \in \mathbb{R}$. We seek a function $f: \chi \rightarrow \mathbb{R}$ that fits these training points and, ideally, generalizes to new inputs. We select this function from a reproducing kernel Hilbert space (RKHS) $\mathcal{H}$, which is a Hilbert space of functions defined by a symmetric positive-definite kernel function $K: \chi \times \chi \rightarrow \mathbb{R}$.

The learning objective balances two competing goals: minimizing prediction errors on training points and controlling function complexity through regularization via the RKHS norm. Specifically, we seek a predictor $f \in \mathcal{H}$ that minimizes the regularized loss
\begin{equation}
    \label{eq:loss}
    \mathcal{L} = \frac{1}{2}\sum_{\mu=1}^{P} (f(x^\mu) - y^\mu)^2 + \frac{\gamma}{2}\lVert f \rVert^2_{\mathcal{H}},
\end{equation}
where $\gamma > 0$ is the regularization parameter and $\lVert f \rVert^2_{\mathcal{H}}$ is the squared RKHS norm. The first term penalizes prediction errors on the training data, while the second term encourages smoothness by penalizing large RKHS norms.

By the representer theorem, despite $\mathcal{H}$ being potentially infinite-dimensional, the optimal solution $f$ can be expressed as a finite linear combination of kernel functions centered at the training points,
\begin{equation}
    \label{eq:representer}
    f(x) = \sum_{\mu=1}^{P} \lambda^\mu K(x^\mu, x),
\end{equation}
where $\lambda^\mu$ are coefficients to be learned. Due to the quadratic nature of both the loss and regularization terms, these coefficients can be found through convex optimization. In particular, these coefficients minimize
\begin{equation}
    \label{eq:loss_coeff}
    \mathcal{L} = \frac{1}{2}\sum_{\mu=1}^{P} \left[\sum_{\nu=1}^P K(x^\mu, x^\nu) \lambda^\nu - y^\mu\right]^2 + \frac{\gamma}{2}\sum_{\mu,\nu=1}^{P} K(x^\mu, x^\nu) \lambda^\mu \lambda^\nu.
\end{equation}
The optimality condition $d\mathcal{L}/d\lambda^\mu = 0$ yields
\begin{equation}
    \label{eq:opt_cond}
    f(x^\mu) - y^\mu + \gamma \lambda^\mu = 0.
\end{equation}

To analyze this system, we use the kernel's eigendecomposition (also called the Mercer decomposition),
\begin{equation}
    \label{eq:mercer}
    K(x,x') = \sum_{i=1}^{N} \rho_i \phi_i(x)\phi_i(x'),
\end{equation}
where $\phi_i(x)$ are an orthonormal basis of eigenfunctions with respect to the data distribution $p(x)$, and $\rho_i > 0$ are the corresponding eigenvalues. Here, $i=1,\ldots,N$ indexes eigenmodes of the kernel with respect to the data distribution, and $N$ is typically infinite. The eigenfunctions satisfy
\begin{align}
    \label{eq:eig_equations}
    \int_{\chi} dx' p(x') K(x,x') \phi_i(x') &= \rho_i \phi_i(x), \\
    \label{eq:eig_ortho}
    \int_{\chi} dx p(x) \phi_i(x) \phi_j(x) &= \delta_{ij}.
\end{align}
This eigenbasis allows us to express both the predictor $f(x)$ and the target function $y(x)$, which generates the observed targets via $y^\mu = y(x^\mu)$, as
\begin{align}
    \label{eq:f_expansion}
    f(x) &= \sum_{i=1}^{N} w_i \phi_i(x), \\
    \label{eq:y_expansion}
    y(x) &= \sum_{i=1}^{N} a_i \phi_i(x),
\end{align}
where $w_i$ and $a_i$ are expansion coefficients. The predictor coefficients relate to the representer coefficients in \eref{eq:representer} through
\begin{equation}
    \label{eq:w_lambda_relation}
    w_i = \rho_i \sum_{\mu=1}^{P} \lambda^\mu \phi_i(x^\mu).
\end{equation}

To track prediction performance, we define deviation variables measuring both prediction and eigenbasis errors,
\begin{align}
    \label{eq:epsilon_def}
    \varepsilon^\mu &= y^\mu - f(x^\mu), \\
    \label{eq:Delta_def}
    \Delta_i &= a_i - w_i.
\end{align}
From the optimality condition (\eref{eq:opt_cond}), we have
\begin{equation}
    \varepsilon^\mu = \gamma \lambda^\mu. \label{eq:new_eqn_yay}
\end{equation}
These variables let us express the training data-averaged training and test errors as
\begin{align}
    \label{eq:train_error}
    E_\text{train} &= \tavg{\frac{1}{P}\sum_{\mu=1}^P (y^\mu - f(x^\mu))^2} = \tavg{(\varepsilon^\mu)^2}, \\
    \label{eq:test_error}
    E_\text{test} &= \tavg{(y(x^*) - f(x^*))^2} = \sum_{i=1}^{N} \tavg{\Delta_i^2},
\end{align}
where $\tavg{\cdots}$ denotes averaging over the distribution $p(x)$ from which the training data are drawn, and $x^*$ represents a test point drawn from the same distribution.

\subsection{Natural scalings}
\label{subsec:scalings}
We analyze this system in a high-dimensional limit where both $N$ and $P$ approach infinity with their ratio fixed. That is, that the number of modes $N$ needed to diagonalize the kernel with respect to the full data distribution $p(x)$ is on the same order as the number of training data $P$. One way this can arise is if $p(x)$ is supported on a finite set of $N$ points that are uncorrelated or weakly correlated under the kernel, and the training set consists of $P = \mathcal{O}(N)$ randomly subsampled points from this support.

Several quantities must scale appropriately with $P$ (equivalently, with $N$) in this limit.

We assume that both $y(x)$ and $f(x)$ are $\mathcal{O}(1)$. From the expansion of $y(x)$ (\eref{eq:y_expansion}) and eigenfunction orthonormality (\eref{eq:eig_ortho}), we have $\tavg{y^2(x)} = \sum_{i=1}^{N} a_i^2$. For this to remain $\mathcal{O}(1)$, the target-function coefficients must scale as $a_i = \mathcal{O}(1/\sqrt{P})$. Since the optimization aims to set $w_i \approx a_i$, we also have $w_i = \mathcal{O}(1/\sqrt{P})$. We assume that the difference $\Delta_i$ between $a_i$ and $w_i$ has the same scaling as each constituent quantity, and thus $\Delta_i = \mathcal{O}(1/\sqrt{P})$. This implies through \eref{eq:test_error} that $E_\text{test} = \mathcal{O}(1)$.

Similarly, we assume that $\varepsilon^\mu$, given by the difference of $\mathcal{O}(1)$ quantities $f(x^\mu)$ and $y^\mu$, has the same scaling as each constituent quantity, and thus $\varepsilon^\mu = \mathcal{O}(1)$. This implies through \eref{eq:train_error} that $E_\text{train} = \mathcal{O}(1)$. Also, from \eref{eq:new_eqn_yay}, we have $\lambda^\mu = \mathcal{O}(1)$.

Finally, we assume $K(x,x) = \mathcal{O}(1)$. Averaging over $x$ and using the kernel's eigendecomposition (\eref{eq:mercer}), we have $\tavg{K(x,x)} = \sum_{i=1}^N \rho_i$. Setting this quantity to $\mathcal{O}(1)$ requires $\rho_i = \mathcal{O}(1/P)$.

\subsection{Random-eigenfunctions assumption}
\label{subsec:random_eigenfunc}
We denote the kernel eigenfunctions evaluated at the training data by $\phi^\mu_i = \phi_i(x^\mu)$. We assume that these quantities satisfy
\begin{align}
    \label{eq:phi_mean}
    \tavg{\phi_i^\mu } &= 0, \\
    \label{eq:phi_covariance}
    \tavg{\phi_i^\mu \phi_j^\nu} &= \delta^{\mu\nu} \delta_{ij}.
\end{align}
The zero-mean condition (\eref{eq:phi_mean}) assumes symmetry of the eigenfunctions. With this assumption, the second-order statistics (\eref{eq:phi_covariance}) follow from eigenfunction orthonormality (\eref{eq:eig_ortho}).

For the bipartite cavity analysis, we make the stronger assumption that the components $\phi_i^\mu$ are i.i.d., analogous to the i.i.d. assumptions in the perceptron problems. Unlike those cases, we will not need to invoke the central limit theorem; the i.i.d. assumption simply ensures that augmenting the system with a statistically independent eigenmode or datum, the key operations in the cavity method, are possible.

\subsection{Bipartite cavity analysis}
Unlike the perceptron problems, kernel ridge regression lacks the symmetry described in \secref{subsec:symmetry}, and thus there does not exist a naïve mean-field analysis that yields correct results. We therefore proceed directly to the bipartite cavity analysis.

The system exhibits a bipartite structure  (\figref{fig:cavity_fig}, ``Kernel ridge regression'') described by
\begin{align}
    \label{eq:epsilon_system}
    \varepsilon^\mu &= \sum_{i=1}^{N} \phi^\mu_i \Delta_i + I^\mu, \\
    \label{eq:Delta_system}
    \Delta_i &= a_i - \frac{\rho_i}{\gamma}\sum_{\mu=1}^{P} \phi^\mu_i \varepsilon^\mu + I_i,
\end{align}
where $I^\mu$ and $I_i$ are source terms for computing response functions. The structure parallels previous problems: feature variables $\Delta_i$ (eigenmode errors) and datum variables $\varepsilon^\mu$ (prediction errors) interact through quenched random couplings $\phi^\mu_i$.

Although naïve independence assumptions do not yield correct results for the full analysis, they provide useful consistency checks for the scaling behavior established in \secref{subsec:scalings}, given the statistical properties of $\phi^\mu_i$ described in \secref{subsec:random_eigenfunc}. Consider first \eref{eq:epsilon_system}: assuming independence between the quenched disorder $\phi^\mu_i$ and dynamic variables $\Delta_i$, the sum $\sum_{i=1}^N \phi^\mu_i \Delta_i$ comprises $N = \mathcal{O}(P)$ zero-mean terms, each of magnitude $\mathcal{O}(1/\sqrt{P})$. This yields an $\mathcal{O}(1)$ result, consistent with the expectation that $\varepsilon^\mu = \mathcal{O}(1)$. Similarly, for \eref{eq:Delta_system}, the sum $\sum_{\mu=1}^{P} \phi^\mu_i \varepsilon^\mu$ (under the independence assumption) consists of $P$ zero-mean terms of magnitude $\mathcal{O}(1)$, producing an $\mathcal{O}(\sqrt{P})$ result. Multiplication by $\rho_i = \mathcal{O}(1/P)$ then yields an $\mathcal{O}(1/\sqrt{P})$ result, matching the expectation that $\Delta_i = \mathcal{O}(1/\sqrt{P})$.

\subsubsection{Datum cavity}

We begin with unperturbed dynamic variables $\varepsilon^\mu$ and $\Delta_i$, then introduce a new datum variable $\varepsilon^{0}$ with quenched random variables $\phi_i^{0}$ connecting it to the existing feature variables $\Delta_i$. The introduction of $\varepsilon^{0}$ delivers a perturbation to the feature variables given by
\begin{equation}
    \label{eq:delta_Delta}
    \delta \Delta_i = \sum_{j=1}^N \frac{d \Delta_i}{d I_j} \left(-\frac{\rho_j}{\gamma}\phi^{0}_j \varepsilon^{0} \right).
\end{equation}
The bipartite structure implies that the perturbations to other datum variables $\varepsilon^\mu$ with $\mu = 1\ldots, P$ do not enter directly into the expression for $\varepsilon^{0}$, which reads
\begin{align}
    \label{eq:epsilon0_start}
    \varepsilon^{0} &= \sum_{i=1}^{N} \phi_i^{0} \left(\Delta_i + \delta \Delta_i \right) + I^{0} \\
    \label{eq:epsilon0_mid}
    &= \sum_{i=1}^{N} \phi_i^{0} \Delta_i - \sum_{i,j=1}^{N} \frac{d \Delta_i}{d I_j} \frac{\rho_j}{\gamma}\phi^{0}_i \phi^{0}_j \varepsilon^{0} + I^{0} \\
    \label{eq:epsilon0_final}
    &= \eta^{0} + F^{00} \varepsilon^{0} + I^{0},
\end{align}
where we define the cavity field $\eta^{0}$ and self-coupling $F^{00}$ as
\begin{align}
    \label{eq:eta0_def}
    \eta^{0} &= \sum_{i=1}^{N} \phi_i^{0} \Delta_i, \\
    \label{eq:F00_def}
    F^{00} &= -\sum_{i,j=1}^{N} \frac{d \Delta_i}{d I_j} \frac{\rho_j}{\gamma}\phi^{0}_i \phi^{0}_j.
\end{align}
The cavity construction ensures that, in these expressions, \textit{the quenched disorder, $\phi_i^{0}$, and dynamic variables, $\Delta_i$ and ${d \Delta_i}/{d I_j}$, are independent of each other}, allowing for calculation of disorder-averaged statistics of these expressions. The cavity field (which is Gaussian, though we will not use this property) has statistics
\begin{align}
    \tavg{\eta^{0}} &= 0, \\
    \label{eq:eta0_var}
    \tavg{(\eta^{0})^2} &= \sum_{i=1}^{N} \tavg{\Delta_i^2} = E_\text{test}.
\end{align}
The self-coupling is order-one and self-averaging with mean
\begin{equation}
    \tavg{F^{00}} = \frac{1}{\gamma} \sum_{i=1}^{N} \rho_i \tavg{\frac{d\Delta_i }{dI_i}}.
\end{equation}
From \eref{eq:epsilon0_final}, we obtain the response function
\begin{equation}
\label{eq:S_epsilon}
    S^{\varepsilon} = \frac{d \varepsilon^{0}}{dI^{0}} = \frac{\gamma}{\gamma + \sum_{i=1}^{N} \rho_i \tavg{\frac{d\Delta_i}{dI_i}}}.
\end{equation}
With this definition, dropping the source term yields
\begin{equation}
    \label{eq:epsilon0_solution}
    \varepsilon^{0} = S^\varepsilon \eta^{0}.
\end{equation}

\subsubsection{Feature cavity}

For the feature cavity, we introduce a new feature variable $\Delta_0$ with quenched random variables $\phi_0^\mu$ connecting it to the existing datum variables $\varepsilon^\mu$. The introduction of $\Delta_0$ delivers a perturbation to the datum variables given by
\begin{equation}
    \label{eq:delta_epsilon}
    \delta \varepsilon^\mu = \sum_{\nu=1}^P \frac{d \varepsilon^\mu}{d I^\nu} \phi^\nu_0 \Delta_0.
\end{equation}
The bipartite structure implies that the perturbations to other feature variables $\Delta_i$ with $i=  1, \ldots, N$ do not enter directly into the expression for $\Delta_0$, which reads
\begin{align}
    \label{eq:Delta0_start}
    \Delta_0 &= a_0 - \frac{\rho_0}{\gamma}\sum_{\mu=1}^{P} \phi^\mu_0 (\varepsilon^\mu + \delta \varepsilon^\mu) + I_0 \\
    \label{eq:Delta0_mid}
    &= a_0 - \frac{\rho_0}{\gamma}\sum_{\mu=1}^{P} \phi^\mu_0 \varepsilon^\mu - \frac{\rho_0}{\gamma}\sum_{\mu,\nu=1}^{P} \phi^\mu_0 \frac{d \varepsilon^\mu}{d I^\nu} \phi^\nu_0 \Delta_0 + I_0 \\
    \label{eq:Delta0_final}
    &= a_0 - \frac{\rho_0}{\gamma}\eta_0 - F_{00}\Delta_0 + I_0,
\end{align}
where we define the cavity field $\eta_0$ and self-coupling $F_{00}$ as
\begin{align}
    \label{eq:eta0n_def}
    \eta_0 &= \sum_{\mu=1}^{P} \phi^\mu_0 \varepsilon^\mu, \\
    \label{eq:F00n_def}
    F_{00} &= \frac{\rho_0}{\gamma}\sum_{\mu,\nu=1}^{P} \phi^\mu_0 \frac{d \varepsilon^\mu}{d I^\nu} \phi^\nu_0.
\end{align}
The cavity construction ensures that, in these expressions, \textit{the quenched disorder, $\phi^\mu_0$, and dynamic variables, $\varepsilon^\mu$ and ${d \varepsilon^\mu}/{d I^\nu}$, are independent of each other}, allowing for calculation of disorder-averaged statistics of these expressions. The cavity field (which is Gaussian, though we will not use this property) has statistics
\begin{align}
    \label{eq:eta0n_mean}
    \tavg{\eta_0} &= 0, \\
    \label{eq:eta0n_var}
    \tavg{(\eta_0)^2} &= P\tavg{(\varepsilon^\mu)^2} = PE_\text{train}.
\end{align}
The self-coupling is order-one and self-averaging with mean
\begin{equation}
    \label{eq:F00n_mean}
    \tavg{F_{00}} = \frac{\rho_0 P}{\gamma}S^\varepsilon.
\end{equation}
From \eref{eq:Delta0_final}, we obtain the response function
\begin{equation}
    \label{eq:S_Delta}
    S^\Delta_0 = \frac{d\Delta_0}{dI_0} = \frac{\gamma}{\gamma + P\rho_0 S^\varepsilon}.
\end{equation}
Dropping the source term yields
\begin{equation}
    \label{eq:Delta0_solution}
    \Delta_0 = S^\Delta_0\left(a_0 - \frac{\rho_0\eta_0}{\gamma}\right).
\end{equation}


To solve for the generalization error, we first determine $S^\varepsilon$. We define
\begin{equation}
    \label{eq:kappa_def}
    S^\varepsilon = \frac{\gamma}{\kappa},
\end{equation}
where $\kappa$ is a parameter to be determined (not to be confused with the margin in the Gardner and manifold capacity problems). Substituting \eref{eq:S_Delta} into \eref{eq:S_epsilon} yields the self-consistency equation
\begin{equation}
    \label{eq:kappa_eqn}
    \kappa = \gamma + \sum_{i=1}^{N} \frac{\kappa \rho_i}{\kappa + P \rho_i}.
\end{equation}
From Eqs.~\ref{eq:epsilon0_solution} and \ref{eq:eta0_var}, we can relate the training and test errors via
\begin{equation}
    \label{eq:train_test_relation}
    E_\text{train} = \frac{\gamma^2}{\kappa^2} E_\text{test}.
\end{equation}
Finally, we determine the test error $E_\text{test} = \sum_{i=1}^{N} \tavg{\Delta_i^2}$ by squaring and averaging \eref{eq:Delta0_solution},
\begin{align}
    \label{eq:Delta0_sq_start}
    \tavg{\Delta_0^2} &= (S^\Delta_0)^2 a_0^2 + \frac{P}{\gamma^2} (S^\Delta_0)^2 \rho_0^2 (S^\varepsilon)^2 E_\text{test} \\
    \label{eq:Delta0_sq_end}
    &= \frac{\kappa^2 a_0^2}{(\kappa + P \rho_0)^2} + \frac{P \rho_0^2}{(\kappa + P\rho_0)^2} E_\text{test}.
\end{align}
Solving this yields the test error
\begin{align}
    \label{eq:test_error_final}
    E_\text{test} &= \frac{1}{1-A}\sum_{i=1}^{N} \frac{\kappa^2 a_i^2}{(\kappa + P\rho_i)^2}, \\
    \label{eq:A_def}
    A &= \sum_{i=1}^{N} \frac{P\rho_i^2}{(\kappa + P\rho_i)^2}.
\end{align}
Together, Eqs.~\ref{eq:kappa_eqn}, \ref{eq:train_test_relation}, and \ref{eq:test_error_final} provide a complete solution for the training and test errors. These expressions match those derived through other replica, cavity, and random-matrix approaches. The solution depends on four key ingredients: the kernel eigenspectrum $\rho_i$, target function coefficients $a_i$, regularization parameter $\gamma$, and number of training points $P$. This explicit dependence reveals how these factors interact to determine learning performance.

\section{Conclusion}

We have illustrated a simple cavity method for high-dimensional convex learning problems. Analysis of three archetypal problems---perceptron classification of points, perceptron classification of manifolds, and kernel ridge regression---reveals their shared bipartite structure. Despite their apparent differences, each system exhibits feature and datum variables interacting through random couplings, enabling unified analysis through the cavity method. This framework should facilitate rapid exploration of novel high-dimensional convex learning problems.

The analysis naturally extends to more general scenarios. For classification problems, one can incorporate slack variables to allow for soft margins. While this generalization has been analyzed using replicas for both point and manifold classification \cite{cohen2022soft}, the bipartite cavity method may offer a clearer derivation.

The method can accommodate additional structure in data and labels. When the separating hyperplane passes through the origin and the pattern distribution $P(x^\mu_i)$ is symmetric about zero, binary labels $y^\mu \in \{-1, +1\}$ do not affect the calculation, making class imbalance irrelevant \cite{chung2018classification}. However, introducing a bias term or using asymmetric pattern distributions breaks this symmetry and makes class imbalance relevant \cite{gardner1988space}, requiring that the labels are tracked through the calculation. 

For correlated features in point classification, a simple form of correlation arises when $P(x^\mu_i)$ is asymmetric (e.g., delta functions at $\pm 1$ with different weights), yielding nonzero $\tavg{x^\mu_i x^\mu_j}$ for $i \neq j$. Since patterns remain i.i.d., the method we have presented applies straightforwardly by retaining labels and accounting for different statistics in the disorder averaging.

A different correlation structure involves patterns drawn from a multivariate Gaussian with anisotropic covariance $\bm{\Sigma}$. In this case one can rotate the data into the eigenbasis of $\bm{\Sigma}$. The makes the problem equivalent to using features that are statistically independent but not identical, with the variance of the $i$-th feature given by the eigenvalue $\lambda_i$ of $\bm{\Sigma}$. The capacity is equivalent to that for classifying $P$ i.i.d. (equal-variance) points in an effective number of dimensions given by the participation ratio of the spectrum, 
\begin{equation}
    \text{PR} = \frac{\left(\sum_{i=1}^N \lambda_i\right)^2}{ \sum_{i=1}^N \lambda_i^2}.
\end{equation}
More complex structures, such as Gaussian mixtures, would require additional variables for cavity calculations.

For manifold capacity problems, one can analyze cases where manifolds exhibit correlations in their centers or orientations, as has been done using the replica method \cite{wakhloo2023linear}. Again, doing this with the cavity method would require additional variables.

The benefit of convexity is that any solution of the KKT conditions (in classification problems) or zero-gradient conditions (in regression problems) is guaranteed to be a global optimizer. In non-convex settings, such conditions are only necessary but not sufficient for global optimality. In problems with complex energy landscapes characterized by exponentially many hierarchically organized solutions (e.g., spin glasses), replica symmetry breaking or equivalent techniques may be required.

The bipartite cavity framework also enables the analysis of dynamical neural networks. Consider a recurrent network where preactivations $x_i(t)$ evolve according to
\begin{equation}
    \frac{dx_i(t)}{dt} = -x_i(t) + \sum_{j=1}^N J_{ij} \phi(x_j(t)),
\end{equation}
where $J_{ij}$ are synaptic weights and $\phi(x)$ is a nonlinearity. When the weight matrix has a product structure $\bm{J} = \bm{L}\bm{D}\bm{R}^T$ (e.g., modeling low-rank connectivity), with potentially correlated columns in $\bm{L}$ and $\bm{R}$, the system can be reformulated as coupled neuronal and latent variables. Recent work has shown how extending the bipartite cavity approach to include two simultaneous cavity variables enables characterization of collective network properties such as dimensionality \cite{clark2024connectivity}.

Recent theoretical work has shown that systems of coupled visible and hidden neurons, with data-determined mutual couplings, provide a unified framework encompassing various associative-memory architectures \cite{krotov2016dense, krotov2020large, ramsauer2020hopfield}. These include classical models like the Hopfield network \cite{amit1985storing}, as well as dense associative memories \cite{krotov2016dense} and modern attention-based architectures \cite{ramsauer2020hopfield}. The bipartite cavity method can be applied to study the dynamics of such systems \cite{clark2025transient}. 

\section*{Acknowledgements}
We thank L. Abbott and A. Litwin-Kumar for comments on the manuscript, and O. Marschall for discussions. We thank A. M. Sengupta for a \href{https://www.condmatjclub.org/jccm_january_2025_03}{commentary on this manuscript}, as part of the Journal Club for Condensed Matter Physics, that pointed out prior work on the zero-temperature cavity method applied to bipartite structures.

\paragraph{Funding information}
D.G.C. was supported by the Kavli Foundation and the Gatsby Charitable Foundation (GAT3708). H.S. was supported by the Swartz Foundation, the Gatsby Charitable Foundation, the Kempner Institute for the Study of Natural and Artificial Intelligence at Harvard University, and the Office of Naval Research (N0014-23-1-2051).

\begin{appendix}
\numberwithin{equation}{section}

\section{Additional manifold capacity analysis}
\label{sec:appendix_manifolds}

\subsection{Computing the response functions}

To complete our analysis of the manifold capacity problem, we compute the response functions explicitly. We begin by expressing
\begin{equation}
    \lambda^{0} \bm{s}^{0} = \frac{\bm{V}^{0} - \bm{T}^{0} - \bm{I}^{0}}{S^w}.
\end{equation}
Differentiating with respect to the source term gives
\begin{equation}
    \frac{d(\lambda^{0}\bm{s}^{0})}{d\bm{I}^{0}} = \frac{1}{S^w}\left[\frac{d\bm{V}^{0}}{d\bm{I}^{0}} - \bm{\mathcal{I}} \right],
\end{equation}
reducing our task to computing ${d\bm{V}^{0}}/{d\bm{I}^{0}}$.

The response behavior differs for supporting and non-supporting manifolds. For non-supporting manifolds where $g_S(\bm{V}^{0}) > 1$, we have $\lambda^{0} = 0$ and thus $\bm{V}^{0} = \bm{T}^{0} + \bm{I}^{0}$. This immediately yields ${d\bm{V}^{0}}/{d\bm{I}^{0}} = \bm{\mathcal{I}}$ and ${d(\lambda^{0}\bm{s}^{0})}/{d\bm{I}^{0}} = 0$. For supporting manifolds where $g_S(\bm{V}^{0}) = 1$, the calculation is more involved. Differentiating the equation for $\bm{V}^{0}$ gives
\begin{equation}
    \frac{d\bm{V}^{0}}{d\bm{I}^{0}} = \bm{\mathcal{I}} + S^w \bm{s}^{0} \left(\frac{d\lambda^{0}}{d\bm{I}^{0}}\right)^T + S^w \lambda^{0} \bm{H}^{0} \frac{d\bm{V}^{0}}{d\bm{I}^{0}},
\end{equation}
where $\bm{H}^{0} = \nabla^2 g_S(\bm{V}^{0})$ is the Hessian of the support function. This Hessian captures the local curvature of the manifold at the anchor point. Additionally, differentiating the constraint $g_S(\bm{V}^{0}) = 1$ yields
\begin{equation}
    \left(\frac{d\bm{V}^{0}}{d\bm{I}^{0}}\right)^T \bm{s}^{0} = \bm{0}.
\end{equation}
To solve this system, we introduce the projection operator
\begin{equation}
    \bm{P}^{0} = \left[\bm{\mathcal{I}} - S^w \lambda^{0} \bm{H}^{0}\right]^{-1},
\end{equation}
which is symmetric. This allows us to write
\begin{equation}
    \frac{d\bm{V}^{0}}{d\bm{I}^{0}} = \bm{P}^{0}\left[\bm{\mathcal{I}} + S^w \bm{s}^{0} \left(\frac{d\lambda^{0}}{d\bm{I}^{0}}\right)^T\right].
\end{equation}
Using the constraint on $d\bm{V}^{0}/d\bm{I}^{0}$, we solve for $d\lambda^{0}/d\bm{I}^{0}$,
\begin{equation}
    \frac{d\lambda^{0}}{d\bm{I}^{0}} = -\frac{1}{S^w (\bm{s}^{0})^T \bm{P}^{0} \bm{s}^{0}} \bm{P}^{0}\bm{s}^{0}.
\end{equation}
Substituting back yields
\begin{equation}
    \frac{d\bm{V}^{0}}{d\bm{I}^{0}} = \bm{P}^{0} - \frac{1}{(\bm{s}^{0})^T \bm{P}^{0} \bm{s}^{0}} \bm{P}^{0} \bm{s}^{0} (\bm{s}^{0})^T \bm{P}^{0}.
\end{equation}
Finally, for supporting manifolds, the trace of the response function is
\begin{equation}
    \text{tr}\left(\frac{d(\lambda^{0} \bm{s}^{0})}{d\bm{I}^{0}}\right) = \frac{1}{S^w}\left[\text{tr}(\bm{P}^{0} - \bm{\mathcal{I}}) - \frac{(\bm{s}^{0})^T (\bm{P}^{0})^2 \bm{s}^{0}}{(\bm{s}^{0})^T \bm{P}^{0} \bm{s}^{0}}\right].
\end{equation}
We unify the supporting and non-supporting cases by defining
\begin{equation}
    \psi^0 = \begin{cases}
        0 & \text{non-supporting},\\
        \text{tr}(\bm{P}^{0} - \bm{\mathcal{I}}) - \frac{(\bm{s}^{0})^T (\bm{P}^{0})^2 \bm{s}^{0}}{(\bm{s}^{0})^T \bm{P}^{0} \bm{s}^{0}} & \text{supporting}        
    \end{cases},
\end{equation}
which allows us to write the self-consistency equation for the weight response function as
\begin{equation}
    S^w = 1 + \alpha \tavg{\psi^0}. 
\end{equation}

\subsection{Resolving the inconsistency}

As in the Gardner problem, we encounter an apparent inconsistency between two expressions for the inverse capacity. The first expression, derived from weight-vector normalization without averaging, is $\alpha^{-1} = \kappa^2 \tavg{\lambda^\mu}$. The second comes from the cavity analysis,
$
    \alpha^{-1} = \kappa^2(S^w)^2 \tavg{\lVert\lambda^\mu \bm{s}^\mu\rVert_2^2}.
$
To check that these are consistent, we first express $\lambda^{0}$ using the KKT conditions as
\begin{equation}
    \lambda^{0} = \frac{h^{0} + 1 - (\bm{T}^{0})^T \bm{s}^{0}}{S^w \lVert \bm{s}^{0}\rVert_2^2}
    = \frac{\left[1 - (\bm{T}^{0})^T \bm{s}^{0}\right]_+}{S^w \lVert \bm{s}^{0}\rVert_2^2}.
\end{equation}
Substituting this into the first capacity formula and using $S^w = 1 + \alpha\tavg{\psi^0}$ yields
\begin{equation}
    \alpha^{-1} = \kappa^2 \tavg{\frac{\left[1 - (\bm{T}^{0})^T \bm{s}^{0}\right]_+}{ \lVert \bm{s}^{0}\rVert_2^2}} - \tavg{\psi^0}.
\end{equation}
Consistency requires the identity
\begin{equation}
    \kappa^2  \tavg{\frac{\left[1 - (\bm{T}^{0})^T \bm{s}^{0}\right]_+^2}{ \lVert \bm{s}^{0}\rVert_2^2}} 
    = - \tavg{\psi^0} + \kappa^2 \tavg{\frac{\left[1 - (\bm{T}^{0})^T \bm{s}^{0}\right]_+}{ \lVert \bm{s}^{0}\rVert_2^2}}.
\end{equation}
We can prove this identity through integration by parts. For clarity, we temporarily drop the $0$ superscripts. The left-hand side can be written as
\begin{equation}
    \kappa^2 \int  \mathcal{D} \bm{T} \left( 1 - \bm{T}^T \bm{s}\right) S^w \lambda 
    = \kappa^2 S^w \tavg{\lambda} - \kappa^2 S^w \int  \mathcal{D} \bm{T} \bm{T}^T \lambda\bm{s},
\end{equation}
where $\mathcal{D}\bm{T}$ denotes Gaussian measure, $\mathcal{D}\bm{T} = d\bm{T} f(\bm{T})$, where
\begin{equation}
    f(\bm{T}) = \left(\frac{\kappa^2}{2\pi}\right)^{(D+1)/2} \exp\left(-\frac{\kappa^2}{2} \lVert\bm{T}\rVert_2^2\right).
\end{equation}
The second term can be evaluated by integration by parts,
\begin{align}
    - \kappa^2 S^w \int  \mathcal{D} \bm{T} \bm{T}^T \lambda\bm{s} &= S^w \int  d\bm{T} \: \left[\grad_{\bm{T}} f(\bm{T})\right]^T \lambda\bm{s}  \\
    &= -S^w \int d\bm{T} \: f(\bm{T}) \grad_{\bm{T}} \cdot (\lambda \bm{s}).
\end{align}
The divergence is simply
\begin{equation}
    \grad_{\bm{T}} \cdot (\lambda \bm{s}) = \text{tr}\left( \frac{d(\lambda \bm{s})}{d\bm{I}} \right),
\end{equation}
completing the proof of the identity.

\end{appendix}





\bibliography{refs.bib}


\end{document}